# Reconfigurable Silicon Photonics Extreme Learning Machine with Random Non-linearities as Neural Processor and Physical Unclonable Function

George Sarantoglou[1,2], Georgios Aias Karydis[3], Adonis Bogris[3], Charis Mesaritakis[1]

1) Department of Biomedical Engineering, University of West Attica, Agiou Spiridonos,12243, Egaleo, Greece.

2) Dept. Information and Communication Systems Engineering, University of the Aegean, Palama 2 str. Karlovasi Samos 83200-Greece.

3) Department of Informatics & Computer Engineering, University of West Attica, Agiou Spiridonos 12243, Egaleo, Athens, Greece.

Corresponding Author: cmesar@uniwa.gr

## Abstract

An alternative extreme learning machine (ELM) paradigm is presented exploiting random non-linearities (RN), named RN-ELM, instead of a conventional fixed node non-linearity. This method is implemented on a hybrid neural engine, with the physical layer realized by an integrated silicon photonic mesh and the digital layer by a simple regression algorithm. Non-linearities are intrinsically non-power depended and are generated through non-linear frequency to power mapping offered by optical filters. The numerical evaluation is based on an experimentally derived transfer function of an all-pass filter, implemented on a silicon reconfigurable photonic integrated chip (RPIC). RN-ELM is evaluated in a twofold manner; first as a machine learning scheme, where the expressivity offered by multiple, yet random, activation functions lead to a compact and highly simplified design with 5 optical filters, offering state-of-the-art performance in time-series prediction tasks with minimum hardware requirements. The second scenario entails its deployment as a physical unclonable function (PUF), for authentication applications directly in the physical layer. In this case, the random activation functions are associated with unavoidable, fabrication related waveguide imperfections that can act as hardware "signatures". Numerical results reveal a probability of cloning as low as $10^{-15}$, which corresponds to a highly secure authentication token.

# 1 Introduction

Over the last decade, Artificial Neural Networks (ANNs) have emerged as a highly efficient paradigm for a wide range of applications, such as image recognition, natural language processing and time-series prediction[1]. ANNs uncover hidden patterns in data, by performing cascaded linear and non-linear transformations guided by the training procedure. When realized in Von Neumann restricted platforms, the growing complexity of ANNs leads to an exponential increase of power requirements - double every 3.5 months - thus raising serious technological and environmental concerns [2]. In response, silicon photonic accelerators in conjunction with digital ANNs have emerged as a promising alternative. By enabling in-memory computing, ultra-fast processing speeds and parallelism via wavelength division multiplexing (WDM), photonics can offer orders of magnitude improvements in computational efficiency [3]. Furthermore, recent advantages in photonic fabrication allows for optimized co-integration with electronics and unlock deployment at the edge [4–6].

Under the paradigm of conventional multi-layer ANNs, the benefits of photonic accelerators have been associated mainly with power efficient linear calculations[3]. However, the absence of a cost-effective, low power and well-defined photonic node activation function remains a substantial obstacle[3]. Activation functions are implemented either via non-linear light-matter interactions or via opto-electronic (O/E) conversions. Kerr effect is the most prominent non-linear light-matter interaction mechanism, but demands high input power and lacks configurability[7]. Opto-electronic (O/E) activation functions provide non-linear responses at substantially lower optical input powers [4, 8], yet scaling is hampered by current high-speed modulation technology, which demands either large physical footprint (lithium niobate, barium titanite oxide) or high optical losses (silicon carrier depletion modulators) along with complex circuit packaging[9].

Another roadblock for photonic accelerators is the lack of an efficient on-chip training technique. Gradient based methods like back-propagation are hard to realize when analog activation functions are considered, whereas alternative methods such as direct feedback alignment[10] and forward-forward training[11, 12] are still immature lagging in terms of performance. To bypass these challenges, bio-inspired algorithms based on random architectures that allow untrained physical nodes, such as reservoir computer (RC) and extreme learning machines (ELM) have been developed [13]. RCs and ELMs are two-layer networks, where the first layer - known as the reservoir - projects input data to a higher dimensional plane via fixed, untrained and random synapses that connect physical neural nodes, sharing the same fixed activation function. Training is restricted to the second layer, whose inputs are driven to a linear regression algorithm realized on a low-power digital module. The difference between RCs and ELMs is that RCs possess cyclic connectivity in the reservoir, which enables physical memory [14]. ELMs being memory-less require tapping at the back-end via digital post-processing. In RC/ELMs, reservoir layers require significant physical overhead, since dimensionality expansion is implemented via multiple hardware synapses and non-linear nodes [15, 16]. Time multiplexed methods have been proposed as an alternative, requiring a single physical node and thus offering the simplest possible setup [17]. However, extra digital post-processing known as masking is required for data projection, whereas time-multiplexing results into a speed penalty, thus excluding high-speed applications.

In this work, we propose a novel CMOS compatible photonic computing paradigm, named random non-linearity based extreme learning machine (RN-ELM), which is based on a variation of the conventional reservoir architecture, while still restricting training to the digital regression layer. The reservoir of the RN-ELM consists of passive optical filters, operating as neural nodes whose activation functions are determined via the non-linear frequency to power (FTP) mechanism[18]. The form of the activation function depends on the frequency detuning between the carrier and the resonance of each filter, thus providing different non-linearities for randomly detuned optical filters and power-independent non-linear transformations[18]. As a result, the neural nodes of the RN-ELM are not restricted to a universal fixed activation function, but on the contrary leverage the use of differentiated, FTP assisted, random non-linearities. Contrary to conventional RCs and ANNs, non-uniform non-linear neurons provide increased expressivity, yielding simplified architectures. Notably photonic implementations based on trainable polynomial networks [19, 20] and Kolmogorov Arnold networks[21] provide increased computational performance with a single layer, compared to multi-layer approaches. Additionally, the RN-ELM concept attains low physical footprint and complexity, since it is inspired by recent advances in non-linear networks with non-uniform, distinct and trainable activation functions[22, 23]. This principle dramatically decreases reservoir complexity by eliminating linear synaptic connectivity, without introducing any speed penalty.

The capabilities of the proposed scheme are demonstrated using experimentally derived non-linear activation functions by a silicon on insulator reconfigurable photonic integrated circuit based on Mach-Zehnder Interferometer (MZI) meshes [24]. These experimental responses are included in a simulated RN-ELM model that targets the Santa-Fe chaotic prediction task as benchmark. The RN-ELM achieves state-of-the art performance[25] with normalized mean square error (NMSE) of 0.053, while utilizing only 5 physical photonic nodes.

An additional key observation is that the frequency detuning values between a fixed laser source and on-chip filters is determined by waveguide imperfections (roughness) that result to random refractive index variations. Thus, RN-ELM non-linearity randomness is linked to unpredictable fabrication related imperfections rendering RN-ELM suitable also as a physical unclonable function (PUF) similar to [26]. In contrast to traditional approaches where authentication keys are stored in non-volatile digital memories susceptible to hardware attacks, PUFs leverage the underlying unavoidable fabrication imperfections to produce secure keys, via challenge-response pair (CRP) interrogation. Thus, the key is hidden in the physical signature of the device. Recently, dedicated silicon photonic PUFs have emerged [27, 28] as a promising security platform, but they involve additional electronic circuits, limiting their practicality. In our previous works on RC-based PUFs [29–31], the underlying waveguide imperfections were mapped to binary keys by post-processing the trained non-linear weights, thus seamlessly combining neural processing and security. Additionally, since enhanced regularization leads to higher stability of the weights (and therefore of the physical key) to noise, here ridge regression is used similar to our previous work[26]. According to these principles, by using the Santa-Fe benchmark as a challenge, the PUF mechanism is evaluated on the RN-ELM, which generates security keys and a probability of cloning equal to $10^{-15}$, matching state-of-art CMOS PUFs. Overall, the proposed RN-ELM combines the inherent complexity of optics and the miniaturization and reconfiguration capabilities of electronics, addressing at the same time data-processing and key generation.

## 2 EXTREME LEARNING MACHINE WITH HETEROGENEOUS NONLINEARITIES

A generalized linear model can be applied on a regression task according to the relation [32]:

$$\hat{y} = f(x) = W \mathrm{G}(x) = \sum_{k=1}^{K} w_k g_k(x) \quad (1)$$

Here, $x \in \mathcal{R}^L$ is a feature vector, $\mathrm{G}(x) = [g_1(x), g_2(x), \ldots, g_K(x)]^T : \mathcal{R}^L \to \mathcal{R}^K$ is a vector of non-linear functions, $W \in \mathcal{R}^{1 \times K}$ the weights of the regression and $\hat{y} \in \mathcal{R}$ is the predicted output. In conventional ELMs, $G(x)$ is data-independent and equal to $\sigma(M_{K \times L} x)$, where $M_{K \times L}$ is a matrix of fixed and random weights and $\sigma(\cdot)$ is an activation function [33]. Given $G(x)$, the weights can be determined by the linear regression algorithm. In the case of RN-ELM, the $G(x)$ vector is defined directly by distinct, random and fixed non-linearities, thus eliminating the need for masking.

The proposed photonic RN-ELM is presented in Fig.1.a, whereas the mechanism for the generation of multiple heterogeneous non-linearities is shown in Fig.1-b. A laser of power $P_{in}$ is frequency modulated (FM) e.g. via its DBR section [34], resulting in a time varied optical frequency $f_{opt}(t) = f_c + f(t)$. The modulated signal is driven to the physical layer comprised of multiple frequency detuned optical filters with transfer function $H_i(f + \Delta f_i)$ for the $i$-th filter. Here $\phi_i = 2\pi \Delta f_i T$ is the applied phase shift at the cavity of the filter, $T$ is its characteristic time and $\Delta f_i$ is the resulting frequency shift. In this work, filters are realized as all-pass micro-ring resonators (MRR), but in general any filter configuration can be used. The frequency to power (FTP) response of an MRR $|H(f)|^2$ has an intrinsically non-linear Lorentzian shape[35]. By applying a phase shift at each MRR filter, the output power is given by

$$P_i(t) = P_{in}|H(f(t) + \Delta f_i)|^2 \ , i = 1,2,...,N \quad (2)$$

Each phase shift $\Delta f_i$ results in a different non-linear shape around the modulation of the optical carrier, thus providing a different non-linear response at the output power, as it is shown in Fig. 1b. Since ELM permits random and fixed non-linearities, this process can become truly passive by exploiting inherent refractive index variations during the fabrication process according to $\delta n \sim U(-0.015,0.015)$ [36], thus resulting in passive phase biases that follow a uniform distribution $\phi_i \sim U(0,2\pi), i = 1,2,...,N$, where $N$ is the number of MRRs [24]. Driven by the FTP mechanism, these passive phase variations lead to different activation functions (Fig. 1-b). The power outputs of the filters are read by an array of photodetectors, that constitute the output of the physical layer. After the physical layer, a digital layer receives the digitized power outputs and implements a linear regression layer, which determines the optimum weights for the target task, eventually implementing Eq. (1).

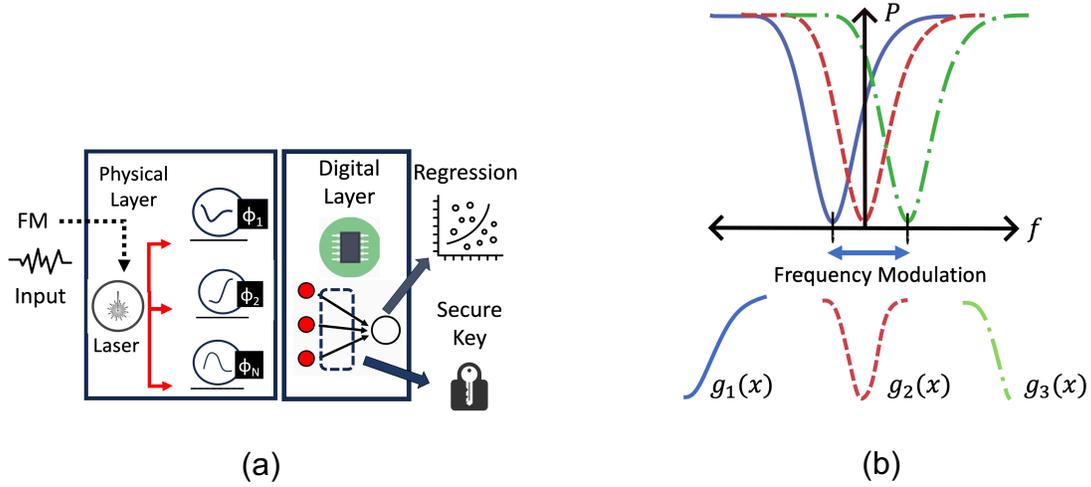

Fig. 1. (a) The RN-ELM concept for regression and key generation. (b) The frequency to power mechanism responsible for the heterogeneous non-linearities.

## 3 EXPERIMENTAL SETUP AND METHODS

### 3.1 Derivation of the transfer function

In order to analyze the RN-ELM concept, the transfer function of an all-pass MRR filter realized using a reconfigurable MZI mesh is derived experimentally. The schematic of the photonic chip is shown in Fig.2-a. It consists of MZIs, called Programmable Unit Cells (PUC)[37] forming hexagonal units. The splitting ratio of each PUC is configured by two thermal actuators. Each PUC can be described by a unitary matrix [38]:

$$U_{2\times2} = je^{j\theta/2}\begin{bmatrix}\sin(\theta/2) & \cos(\theta/2) \\ \cos(\theta/2) & -\sin(\theta/2)\end{bmatrix} \quad (3)$$

Here, $\theta$ is the phase difference programmed by the thermal heaters. The transmissivity and reflectivity of each PUC are defined as $r = \cos^2(\theta/2), t = 1 - r = \sin^2(\theta/2)$, which dictate its splitting ratio. If $r = 1, t = 0$, then the PUC is at bar state as it is the case for PUCs 1-7 in Fig. 2.a. The PUC 8 is configured so as to achieve the critical coupling condition. Each PUC has a length equal to 730 µm. Therefore, the hexagon in Fig. 2.a. forms an MRR with cavity length $L = 6 \times 730 \mu m = 4.38\ mm$. The laser source is set at 8.8 dBm power and its frequency is swept from 193.39 THz to 193.6 THz with a step size of 0.02 GHz. The acquired transfer function is shown in Fig.2-b. The optical bandwidth of the MRR is equal to 700 MHz, the FSR is $\Delta f_{FSR} = 20$ GHz and the extinction ratio is 20 dB. The peak power is $-15.2$ dBm owing to the 24 dB insertion losses of the MRR. This elevated value is the product of accumulated PUC's insertion losses that are tuned to realize the MRR.

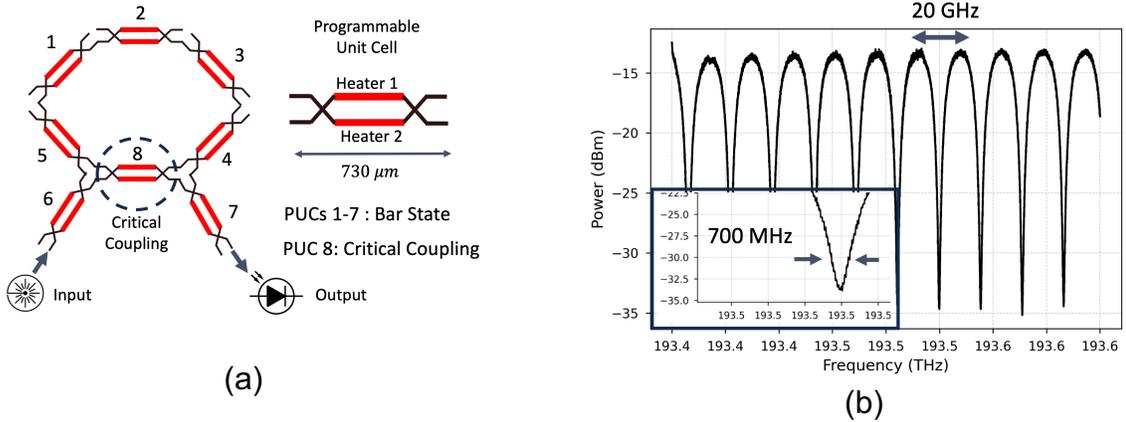

Fig. 2. (a) The RPIC programmed as an all-pass MRR. (b) The transfer function of the implemented filter.

## 3.2 Simulation Based on Experimental Transfer Function

The experimentally derived transfer function (Fig. 2.b) is employed in a simulation environment to benchmark the RN-ELM targeting the Santa Fe chaotic time-series prediction task[39]. The objective is to use past samples of the signal to predict the next sample. The training and testing datasets consist of 1000 samples each. In order to simulate the RN-ELM using the experimentally derived transfer function, the acquired power spectrum was raised at -1.48 dBm per filter, assuming a non-reconfigurable chip with lower propagation losses. For information encoding, FM is used by modulating the DBR section of a laser with central wavelength of 1550 nm. The frequency to power mechanism is simulated by applying an interpolation function $P = z_{int}(f)$ at the transfer function and varying the frequency according to the normalized input signal $n(t)$, as $f_s(t) = s\Delta f_{FSR}\, n(t)$. Here, $s$ is a scaling hyper-parameter that requires optimization. Low values of $s$ lead to a small modulation depth and linear response, whereas high $s$ values result in highly non-linear responses, which due to the periodicity of the transfer function have detrimental effects by mapping different inputs to the same outputs (information aliasing). The optimum value of $s$ is acquired for the Santa Fe task via parametric sweeping. This simulation method ignores the transient effects present in the MRR and laser cavities. For the MRR the photon lifetime is given by $\tau_{ph} = 1/(2\pi\delta v_c) = 220\ ps$. For the DBR laser modulation bandwidths up to 10 GHz have been reported[40]. Therefore, a data rate of 1 Gsa/s is chosen, where the signal variations can be considered much slower compared to the MRR and laser dynamics, allowing for transient effect to be safely ignored. Finally, for the encoding phase, 8-bit quantization is applied to simulate the limited precision of the signal source.

The input is driven into $N$ optical filters. For each filter, a random passive offset $\phi_i \sim U(0,2\pi)$ is defined, which results in a frequency offset $f_i = \phi_i/(2\pi T)$, where $T = 1/\Delta f_{FSR}$ is the round-trip time of the MRR. Therefore, for each filter the evaluated output is $P_i(t) = z_{int}(f_i + f_s(t))$. To simulate the effect of noise, each output $P_i(t)$ is transformed to an optical field with phase noise, according to the laser rate equations-[41]. This field is driven to a photodiode with simulated shot and thermal noise and bandwidth equal to 1 GHz, implemented with a 4th order Butterworth filter. The number of photodiodes is equal to the number of filters, whereas their outputs are the photocurrent traces $i_i(t)$.

At the digital stage, the photocurrent is sampled and the values are normalized. Since the photonic platform is memoryless, tapping is required with a memory of $N_T$. Thus, at each instance, the feature vector consists of $KN_T$ elements, containing the $K$ outputs from all $N_T$ previous time steps. The resulting input vectors are driven to the regression algorithm. The total number of weights is equal to the length of the input vector $N_W = KN_T$.

## 3.3 Physical Unclonable Function

The non-linearities are defined by the random passive offsets that arise from the fabrication process. Therefore, each different device holds a unique randomly sampled $\phi$ −vector, which constitutes its signature. Applying the Santa-Fe benchmark on different physical layers – different chips – provides distinct power responses, which in turn lead to unique digital weights. These weights can be binarized to generate a secret key per device for device authentication. In this process, regularization has the effect of minimizing noise-induced overfitting, thus stabilizing the weights and producing noise-robust keys. For this reason, Bayesian ridge regression (BRR) is used with a Gamma prior distribution, since it provides strong regularization without sacrificing performance[42]. This technique has not been followed in our previous work and it offers a new vista to the dual use photonic integrated approach.

To generate each secret key, each weight-vector $w \in \mathcal{R}^{N_W}$ is projected at a higher dimensional plane via the matrix $\Theta = SFV$, where $V = -diag(\{-1,1\}^{N_W}) \in \mathcal{R}^{N_W \times N_W}$ with probability $\Pr[V_{ii} = \pm 1] = 0.5$, $F \in \mathcal{R}^{N_W \times N_W}$ is the discrete Fourier transform and $S \in \mathcal{R}^{M \times N_W}$ is a sparse matrix with $M = 1000$ elements drawn by a uniform distribution $U(0, N_W)$[43]. The derived vector $\bar{w} \in \mathcal{R}^M$ is uniformly spread using the cumulative distribution[31] and binarized with 1–8-bit precision and Gray encoding. Last, a random sub-set of 256-bits is sampled from the binary vector to form the key-string $b \in \mathcal{R}^{256}$, where $b$ holds only binary values.

## 4 RESULTS

### 4.1 Heterogeneous Non-linearities

Heterogeneous non-linearities are presented in Fig. 3 for $K = 7$ filters and $s = 0.5$. The phase offsets are randomly drawn by $\phi_i \sim U(0, 2\pi)$. Each filter has a unique offset compared to the central carrier owing to its physical fingerprint. Different passive offsets yield different non-linear activation functions at the output. Obviously, randomly biasing of the phase offsets of the MRRs, the non-linear responses present low variability, owing to the periodicity of the filters. In order to enhance the variability of the non-linear functions more complex transfer functions can be deployed by more advanced designs like coupled resonators optical waveguides, multi-stage filters and feedback loops [35].

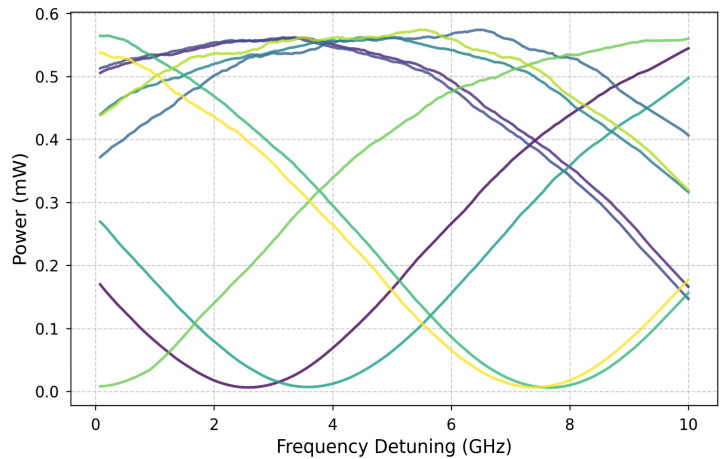

Fig. 3. The non-linear frequency to power mapping. Seven random non-linear functions based on distinct passive offsets.

## 4.2 Regression Performance

The computational performance of the RN-ELM is determined by the NMSE metric, evaluated for the testing set of the Santa-`Fe benchmark. The linear shift-register acts as the baseline with an NMSE of 0.17, which saturates at $N_T = 21$ taps. The RN-ELM is evaluated in terms of the number of non-linear functions $K$. By parametric sweep, the optimum $s$-parameter is $s = 0.6$, whereas $N_T = 11$, since additional taps return non-significant performance improvements. For each $K$ value, the RN-ELM is evaluated 100 times with different $\Phi = (\phi_1, \phi_2, \ldots, \phi_K) \in \mathcal{R}^K$ random sets to determine the average NMSE and its standard deviation.

Results are shown in Fig.4. It can be seen that as $K$ is increased, the average NMSE and its standard deviation are both reduced. For $K = 5$, the NMSE is equal to 0.053. To highlight the simplicity of the RN-ELM scheme, state-of-the-art [25] on-chip performance of NMSE 0.06 has been reported for the same task (both in experiment and simulation) by a spatial RC of higher complexity, consisting of 16 delay lines, a $16 \times 32$ sparse matrix array with 31 MZIs for masking and 32 coherent cavities with 16 virtual nodes each, totaling to 512 nodes [15]. Performance stagnates for higher $K$ values due to overfitting [32], which can be reduced by configuring the non-linearities according to the training data [32] via on-chip [4] or off-chip [6] training. In the suggested setup, this can be achieved by optimizing the phase settings per filter. It should be highlighted that an increase in computational density can be achieved by utilizing WDM via multiple FM modulated DBR lasers. Thus, by combining FM with multiple filters, various complex non-linearities are acquired at a low cost (passive physical layer), thus enabling a highly compact design along with adequate performance.

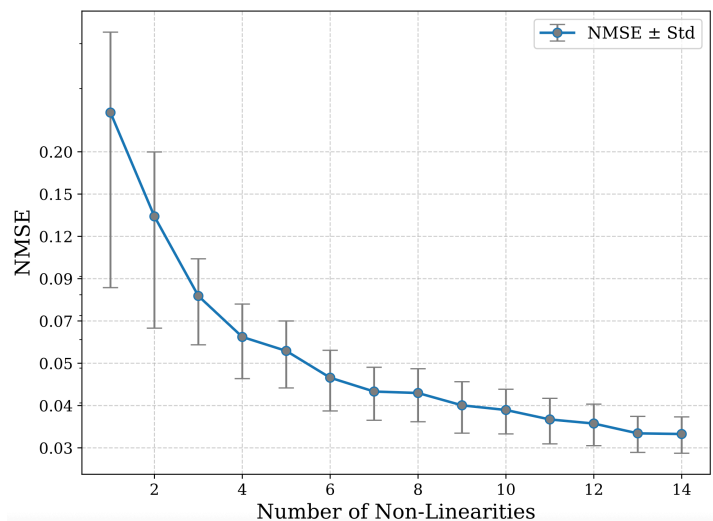

Fig. 4. NMSE as a function of the number of non-linearities. Here $s = 0.6, N_T = 11$.

## 4.3 Physical Unclonable Function

For the PUF scenario, the parameters $K = 9$ and $N_T = 11$ are considered resulting at an NMSE equal to 0.04. Therefore, the total number of weights at the regression layer is equal to 99, which are post-processed according to the presented random projection algorithm to develop keystrings. A bit precision of 3 bits is assumed except if stated otherwise. The two important characteristics of a PUF are the robustness to noise and un-predictability. In order to evaluate these attributes, the normalized Hamming distance (NHD) is defined as the keystring distance metric. NHD among two bit-sequences is defined as $NHD = N_F/256$, where $N_F$ is the total number of bit-flips between them.

A PUF is robust, when its interrogation with the same challenge produces key-strings with low NHD. To statistically determine robustness, 20 devices are considered and 60 key-strings are generated per device considering photodetection noise. The NHD among all 60 key-strings is calculated, leading to 1770 samples per device, totaling to $20 \times 1770 = 35400$ samples. The resulting distribution is known as the intra-distribution and it is depicted as a histogram in Fig. 5. It is fitted with a mixture of two Gaussian probability density functions (PDF), with mean NHD value of 0.0162 and full width at half maximum (FWHM) of 0.03.

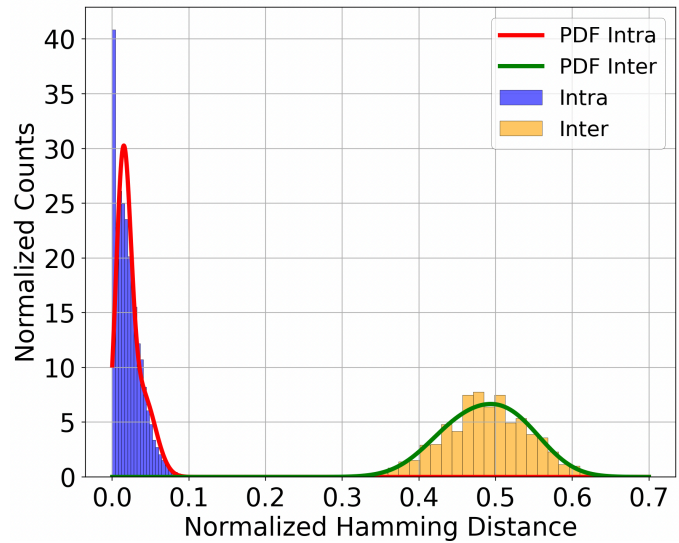

Fig. 5. The Intra (robustness) and Inter distributions (un-predictability) as a function of the NHD. PFDs are fitted with mixtures of two Gaussians. The setup consists of 9 non-linear functions, 11 taps are used and weights are binarized with 3-bit precision.

On the other hand, a PUF is un-predictable if the key-strings produced are highly uncorrelated. To examine this attribute, 1000 different devices are evaluated to produce key-strings, with each device corresponding to a different vector of passive offsets $\Phi \in \mathcal{R}^K$. The NHD among all generated keystrings is calculated, resulting at 499500 samples. This distribution is known as the inter-distribution and its mean value ideally is ideally close to 0.5 (true randomness). The distribution is presented by a histogram in Fig. 5. It is fitted with a mixture of two Gaussians PDF, who's mean NHD is 0.49 and the FWHM is 0.13.

Given the derived robustness and un-predictability PDFs, a NHD value $d_T$ is defined by the authenticator, that acts as a decision threshold. If the NHD between the received and the expected response is higher than $d_T$, then the connection attempt is rejected, assuming that it originates from a difference device. If the NHD is lower than $d_T$ then it is accepted, with the difference from the expected key being attributed to noise. It is evident that the choice of $d_T$ depends on whether higher robustness (closer to the inter PDF) or higher un-predictability (closer ro the intra PDF) is required. The trade off point between these two cases is the equal error rate (EER) threshold $d_{EER}$, where the false rejection rate (FRR) - the probability that a response generated from the same device is not accepted - is equal to the false acceptance rate (FAR) – the probability that a response generated by another device is accepted. FAR and FRR are computed as $FAR(d_T) = P_{intra}(d \leq d_T)$ and $FRR(d_T) = 1 - P_{inter}(d \leq d_T)$ and the EER is $P_{EER} = FAR(d_{EER}) = FRR(d_{EER})$. An EER lower than $10^{-6}$ is linked to adequate performance, whereas for sensitive applications an EER of $10^{-9}$ is required [26]. By using the fitted PDFs, the EER for Fig. 5 is computed equal to $1.77 \times 10^{-11}$, demonstrating strong performance.

A parametric sweep is performed with respect to the number of non-linearities ranging from 1 to 19 and the bit precision from 1 to 8 bits, so as to examine the effect of these parameters on the EER. The results are shown in Fig. 6. By increasing the number of filters, the EER decreases owing to the higher complexity of the physical layer. EER values lower than $10^{-9}$ are observed for $K \geq 9$, whereas for 18 functions it can reach as low as $10^{-15}$. The optimum bit-precision is found at 3 bits. Lower values have the effect of lowering variations since fewer quantization levels are used. This increases robustness, but at the same time decreases unpredictability. On the other hand, higher bit precision values utilize more quantization levels, thus increasing variability, which has a detrimental effect to the robustness, since noise-induced variations appear as bit-flips.

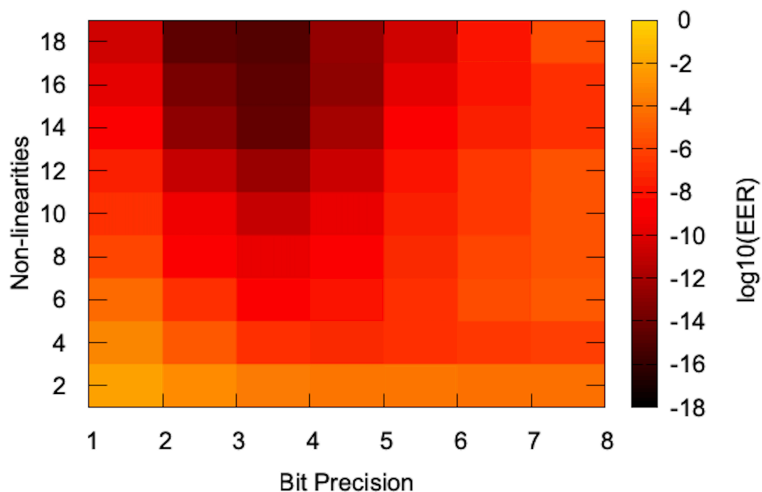

Fig. 6. The EER as a function of the bit-precision and the number of non-linearities. Parameters $s = 0.6, N_T = 11$ are considered.

## 4.4 The effect of Signal to Noise Ratio

Up to this point, a data rate equal to 1 Gsa/s has been considered with an average SNR equal to 35 dB dictated by the thermal noise, which explains the excellent performance of the setup. To study the limitations imposed by the SNR with respect to the NMSE and EER, the system with 9 non-linearities and 3-bit precision is chosen and the input power is varied to produce different SNR levels ranging between 0 and 35 dB. The results are shown in Fig. 7. For the NMSE, an SNR beyond 10 dB is required for performance lower than 0.06, whereas a plateau is reached for SNR values higher than 15 dB. EER is more sensitive to the SNR, with values higher than 30 dB being required for excellent performance – EER lower than $10^{-9}$. Acceptable performance – EER lower than $10^{-6}$ – is available for SNR values higher than 23 dB. The degradation of the PUF operation for decreased SNR is linked to the increase of the mean NHD for the intra-distribution due to the enhanced noise-induced bit-flips.

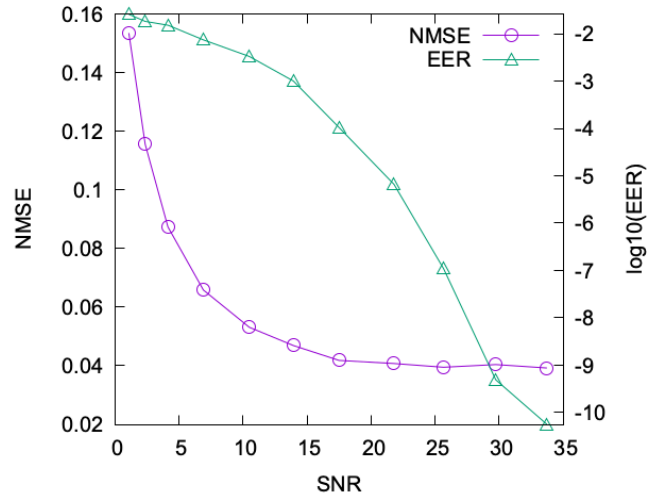

Fig. 7. The NMSR and EER as a function of the SNR. The parameters $K = 7, s = 0.6, N_T = 11$ are considered.

## 5 DISCUSSION AND CONCLUSION

The RN-ELM is presented at a data rate equal to 1 Gsa/s and it can scale up to 20 Gsa/s given the highest modulation rate of 10 GHz for DBR sections in literature [40]. Since the driving mechanism behind the RN-ELM is the non-linear frequency to power mapping, it can also be achieved by alternative methods that exploit ultra-fast modulators (>50 GHz)[9] based on Lithium Niobate[44], Barium Titanite Oxide [45] and carrier depletion[46, 47]. A possible route is to harness phase modulation due to its close relationship to FM by employing differential phase encoding [18]. Another option is to keep the output of the laser unmodulated and drive it to multiple filters, with each filter having a high-speed phase shifter in its cavity. By phase modulation, the frequency detuning between the resonance of the filter and the input is rapidly changing, leading again to the desired non-linear frequency to amplitude mapping.

In conclusion, the RN-ELM paradigm has been presented along with its implementation in a silicon photonics platform, acting both as a neural engine and a PUF, by leveraging non-linear frequency to power transformations via optical filters. The scheme is hybrid, since it combines a physical untrained silicon photonics layer with a digital regression layer.

Numerical results are based on experimentally retrieved non-linearities via an all-pass MRR filter implemented on a RPIC. The RN-ELM generates in a cost-effective manner (passive physical layer) multiple random yet fixed heterogeneous non-linearities and achieves excellent performance on the Santa-Fe benchmark on par with other RC/ELM in literature. The expressivity of the heterogeneous non-linearities leads to highly compact designs (5 optical filters, no masking), which is crucial for edge applications. Although MRR filters are examined, the RN-ELM paradigm is a general method that can exploit any type of optical filter. At the same time, since the non-linearities are directly linked to unavoidable and unpredictable waveguide variations by the fabrication procedure, the weights produced at the digital layer act as a physical fingerprint and can be post-processed to generate secure keys for authentication. Thus RN-ELM acts also as a PUF module, whose performance is evaluated with a probability of error that can reach down to $10^{-15}$. Therefore, this work paves the way towards compact, cost-effective and computational powerful silicon photonic neural engines merged with an inherent authentication mechanism, thus constituting a highly suitable paradigm for an Internet of Things environment in line with the ICT framework.

## Declarations

**Funding**: This work was funded in part by PROMETHEUS (ID: 101070195), QPIC1550 (ID: 101135785) Horizon Europe projects. G. Sarantoglou is supported by the Project QUASAR which is implemented in the framework of H.F.R.I call "Basic research Financing (Horizontal support of all Sciences)" under the National Recovery and Resilience Plan "Greece 2.0" funded by the EU – Next Generation EU No: 016594.

**Author contributions:** G.S. developed the methodology, implemented the numerical simulations and wrote the manuscript with help from all the authors. G.A.K. performed the experiment and extracted the data that were used by the numerical simulations. C.M. conceived the concept C.M together with A.B orchestrated all the work and data processing.

**Competing interests:** The authors declare no competing interests.

Clinical **Trial Number:** Not applicable

**Ethics, Consent to Participate, and Consent to Publish declarations**: not applicable